# Design of a Programmable Gain Waveform Digitization Instrument for Detector Calibration

Zhe Cao, *Member, IEEE*, Jiadong Hu, Cheng Li, Shubin Liu, *Member, IEEE*, Qi An, *Member, IEEE*

*Abstract*—To test and calibrate various detectors in nuclear and high energy physics experiments, a general purposed calibration instrument has been developed. All information including timing, amplitude and charge of signals can be directly obtained to calibrate detector with this instrument by amplifying and digitizing the signal waveform. The system consists of two parts, a large dynamic range pre-amplifier module and a high speed and high-resolution digitization module. The pre-amplifier module based on programmable gain amplifier and attenuator achieves gain from -26.2 dB to 24.4 dB, making it suitable to adapt to different detectors. Taking advantage of high speed and high-resolution analog-to-digit converter (ADC), the waveform digitization module samples the signal after conditioning with 12 bits resolution and up to 3.6 giga hertz samples per second (GSPS). To evaluate the feature of this instrument, an electronics testbench platform was installed and test results showed a bandwidth from direct current (DC) to 450 MHz and the effective number of bits (ENOB) is above 7 bits in high gain set and 5.8 bits in low gain set in the bandwidth range, which indicated that it had broad application prospect in detector calibration.

## I. Introduction

IN nuclear and high energy physics experiment, detector is the core of the whole system. Nowadays, the hybrid detection method is more and more popular in the modern large physics experiment. More kinds of detectors may get more accurate measurement and more information of the physics event. Additional, for different experiment, the signal of different kinds of detectors have various features in amplitude, rising time, bandwidth. Also, in some particle physics experiments, detector signals have an ultra large dynamic range. For example, the Large High Altitude Air Shower Observatory (LHAASO) is a hybrid extensive air shower array with an area of about 1 km[2] at an altitude of 4410 meters above sea level, which is planned to build in Sichuan province, China, to explore the origin of high-energy cosmic rays [1]. Such a detector array is not only useful for the gamma ray source survey but also plays an important role of bridging between direct measurements of energy spectra of individual cosmic ray species at balloon heights and ultra high energy cosmic ray experiments on the ground. In order to fulfil all the goals, a large scale complex of many kinds of detectors is needed [2]. Covering a total area of 1.2 km2, LHAASO consists of a kilo meter square array (KM2A) which includes with 5242 electromagnetic particle detectors (EDs) and 1171 muon detectors (MDs) [3], a water Cherenkov detector array (WCDA) which covers an area of 78000m[2] and consists of three water ponds that two is 150 m × 150 m and the third one is 300 m × 110 m, partitioned into 3120 detector cells [4], and a wide-field Cherenkov telescope array (WFCTA) which composed of 24 telescopes [5].

As a result, the calibration of the detector is always an important task to make sure that the experiment works well. In the traditional way, it is necessary to build a series of calibration electronics for kinds of detectors, so that each electronics can match the characteristic of the particular detector. It is of high cost, low integration and inconvenient in maintenance and management. In the KM2A of the LHAASO, photomultiplier tubes (PMTs) are need to test the photocathode uniformity, linear dynamic range, time spread and dark pulse rate. The data acquisition of the calibration system consists of a VME crate and a NIM crate including a charge-to-digital converter (QDC) module (V965), a low threshold discriminator (N845), a time-to-digital converter (TDC) module (V775N) and a constant fraction discriminator (N843) [6]. In the WCDA, PMTs require not only good single photoelectron (SPE) resolution and small transit time spread (TTS), but also good linearity up to 4000 photoelectrons [7]. To measure the time and charge of the PMT, a multi-hit TDC (V1290A) and a QDC (V965A) were applied with a 10 times amplification (N979) and a low threshold discriminator (V814) [8].

As mentioned above, the particle is generally measured by time-to-digital and voltage-to-digital, which follows the pre-

Manuscript received June 24, 2018.
This work was supported by the National Natural Science Foundation of China under Grant 11505182, and it was also supported by the Anhui Provincial Natural Science Foundation of China under Grant 1608085QA21.

Zhe Cao is with State Key Laboratory of Particle Detection and Electronics and Department of Modern Physics, University of Science and Technology of China, Hefei, 230026, China (e-mail: caozhe@ustc.edu.cn)
Jiadong Hu is with State Key Laboratory of Particle Detection and Electronics and Department of Modern Physics, University of Science and Technology of China, Hefei, 230026, China (e-mail: hjdemail@mail.ustc.edu.cn)
Cheng Li is with State Key Laboratory of Particle Detection and Electronics and Department of Modern Physics, University of Science and Technology of China, Hefei, 230026, China (e-mail: licheng8@mail.ustc.edu.cn)
Shubin Liu is with State Key Laboratory of Particle Detection and Electronics and Department of Modern Physics, University of Science and Technology of China, Hefei, 230026, China (e-mail: liushb@ustc.edu.cn)
Qi An is with State Key Laboratory of Particle Detection and Electronics and Department of Modern Physics, University of Science and Technology of China, Hefei, 230026, China (e-mail: anqi@ustc.edu.cn)

amplifier, charging integration and shaping circuit, implemented by commercial equipment. In order to obtain more information of the signals, a group of complex circuit is employed. Because of the difference among different detectors, more electronics systems will be employed for the calibration. Compared with the traditional method, waveform digitization that digitizes the entire waveform directly, can significantly reduce the order of circuit complexity. Not only the amplitude and arrival time of the detector signal can be acquired, but also the waveform recognition and signal screening of the particle event can be analyzed after processing the discrete digital sequence of the original analog waveform [9]. Additional, a wide dynamic input range is necessary to cover the kinds of the detectors as more as possible.

In this paper, a calibration instrument for various detectors is proposed. The instrument is based on programmable gain amplifying and waveform digitizing, to meet the dynamic range of the amplitude and bandwidth of different kinds of the detector signals. Feature of acquiring maximum information makes fast digitization has advantages in detector calibration. The system architecture with two parts is descripted in detail while a series of performance tests is shown and the results are discussed.

## II. ELECTRONIC SYSTEM DESIGN

### A. System structure

The instrument consists of two modules, a pre-amplifier module and a waveform digitization module. The pre-amplifier module is applied for analog signal conditioning which comes from detector. The waveform digitization module is used to obtain the whole waveform of the conditioned detector signal from the pre-amplifier module. All information including timing, amplitude and charge of signals can be directly acquired from the digitized waveform.

Because of the different feature of various detector employed in diverse nuclear and high energy experiment, this universal instrument should cover a large dynamic range of amplitude of the input signal and a wide bandwidth as low as DC. It also needs to manage both the unipolar positive or negative signal and the bipolar signal. In order to realize a better performance of timing and amplitude measurement, it needs to have a high sample rate with high resolution. As a result, the instrument is designed to digitize the detector signal with large amplitude dynamic range over 50 dB and the bandwidth from DC to about 400 MHz at least with a sample rate of 3.6 GSPS and resolution of 12 bits.

### B. Pre-amplifier module

The proposed architecture, as well as displayed in Fig. 1, composes four main parts: variable attenuator, variable gain amplifier (VGA), differential amplifier and digital controller.

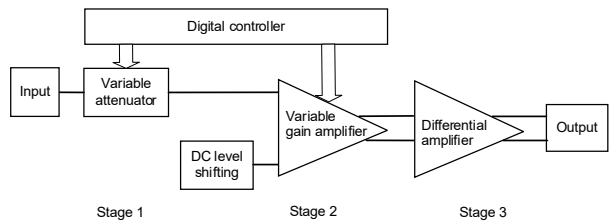

Fig. 1. Schematic of the pre-amplifier module, which has four main components: variable attenuator, VGA, differential amplifier and digital controller.

The bandwidth of the most detectors employed in the nuclear and high energy physics experiments is always as low as DC or ultra low frequency, DC couple is chosen in the module. Besides, to achieve high speed and low distortion, each analog part of this amplifier module should have a wide bandwidth and a low noise.

The first stage is designed to weaken the original detector signal if it is larger than the input range of the following circuit, thus increasing the dynamic range of the system. After compared with single pole double throw switch resistance network attenuation and programmable attenuator chip, a programmable attenuator DAT-31R5-PP+ that has a better frequency response and smaller step is selected. It is a 50 Ohm radio frequency (RF) digital step attenuator that can offer an attenuation range up to 31.5 dB in 0.5 dB digital controlled step with bandwidth from DC to 2400 MHz [10].

VGA is the core component in the module to achieve variable gain of the signal after the programmable attenuator. A digitally controlled VGA AD8370 converts the single-ended signal of the output of the detectors to differential with the desired gain. AD8370 with a DC to 750 MHz bandwidth and 7-bit digitally controlled gain from 6 dB to 34 dB has three internal stages including pre-amplifier, transconductance and output amplifier [11]. The detector signal is fed into the positive input of the chip while the negative input is driven by a digital controlled DC level shifting, which can be adjusted to the required level, comprising of a digital-to-analog converter (DAC).

The amplified signal is then fed into a differential buffer, which can further amplify the detector signal to match the input feature of the high speed analog-to-digital circuit and increase the output drive capability of the module.

An additional controller is designed to fully implement the remote control operation for this module. The control interface of the programmable attenuator 9-bit parallel bus, while DAC and VGA are controlled by serial peripheral interface (SPI) with clock, data and chip select. A microcontroller chip STM32F427 produces all of the remotely digital control interfaces.

The photo of the pre-amplifier module is shown in Fig. 2.

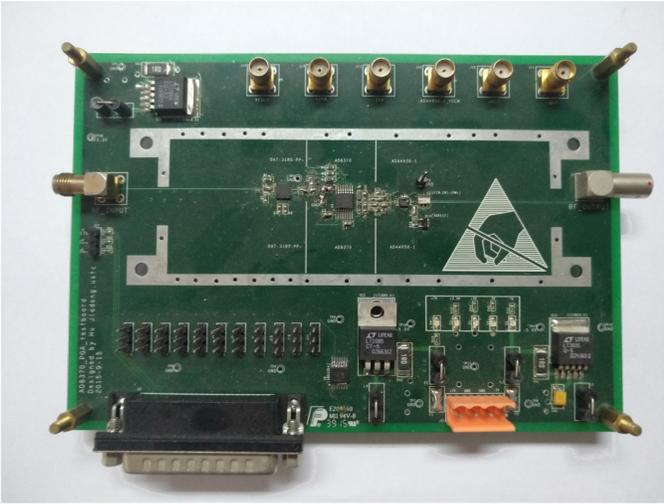

Fig. 2. The photo of the pre-amplifier.

## C. Waveform digitization module

The waveform digitization module is designed as a 3U PCI extension for instrument (PXI) module, as shown in Fig. 3. There are two major types of method to achieve waveform digitization, switched capacitor array (SCA) and high speed and high-resolution ADC. Compared with the benefit and the disadvantage of these two methods, high speed and high-resolution ADC based method has less dead time and longer sample time than SCA based method. One ADC chip, with the sample clock came from a phase locked loop (PLL), is responsible for digitizing the waveform of the conditioned detector signal from the pre-amplifier module. The programmable logic controllers are used to process the digitized data from the ADC and communicate with the PXI controller via PXI crate backplane. Besides, the module can receive the external trigger from the detector system.

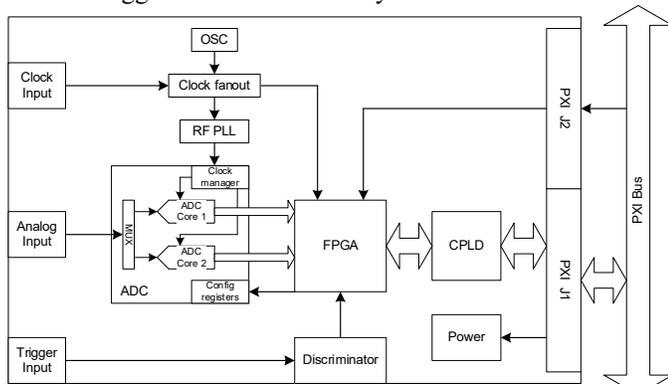

Fig. 3. Schematic of the waveform digitization module, which can digitize the waveform of the detector signal. The main parts of the system consist of ADC, clock system, external trigger receiver, as well as PXI interface and data processing.

According to the requirement of different nuclear and high energy physics experiments, an ADC chip based on folding and interpolating, ADC12D1800 manufactured by TI Company, is chosen to be the critical component of the module. This ADC has two internal 12-bit ADC channels, channel I and channel Q, which can work either independently or in parallel, each with up to 1.8 GSPS sampling rate [12]. In the time interleaved mode, both I and Q channel sample the same analog signal. One channel samples the input on the rising edge of the sampling clock and the other samples the same signal on the falling edge of the sampling clock. A single input is thus sampled twice per clock cycle, resulting in an overall sample rate of twice the sampling clock frequency. With this time interleaved method, the performance of the analog-to-digital system is affected by three main mismatch errors, such as gain error, offset error and time skew error. Digital algorithms are used to correct these errors to improve the performance of the instrument, which have been tested with the mismatch error correction.

In this module, the clock system has an alternative clock source input. There is a clock receiver to receive the external clock so that the module can work with the common clock of other electronics of the experiment. Also, a local high-performance crystal oscillator is set up for the independent operation of the instrument. The high frequency sample clock of the ADC is generated by a RF PLL, LMX2581 manufactured by TI Company, with integrated voltage-controlled oscillator (VCO) [13], which produces a high frequency clock up to 1.8 GHz to sample the input signal in the ADC.

In order to digitize the detector signal in synchronization sample window, external trigger from the detector is an important signal for readout electronics. In this design, two kinds of trigger receiver are considered. In the architecture of the PXI, a star trigger links to all of peripheral slots from star trigger slot. This module can receive the star trigger via backplane if necessary. Additional, it can receive the trigger signal via coaxial cable in front panel. The trigger is discriminated by comparator with preset level, which is set by the DAC, in order to be compatible with different levels of digital signal.

There are two programmable logic devices designed for data processing and interface. The one channel of ADC with 1.8 GSPS and 12 bits channels converts the analog input to 24 bits, 900 Mbps low-voltage differential signaling (LVDS) data streams, while there are 48 pairs of high speed differential transfer lines in total. Artix7 with the benefit of low power consumption and low price is chosen as the FPGA. 48 bits with 900 Mbps data streams are received and demultiplexed to 192 bits with 225 Mbps that can be buffered and reconstructed in the FPGA. It is also used to configure and control all of the components in the module, report the status of the board and find the triggered data. A ring buffer is used to store the data before trigger, and the particular data will be matched according to the trigger latency. Now about 100 us are available because of the limit of internal memory. A MAX II CPLD is performed to be the PXI interface and realize the DMA transfer. The Intellectual Property core of PCI Compiler achieves the PXI bus [14]. The advantage of this design is that the CPLD and firmware can be reused.

The photo of the waveform digitization module is shown in Fig. 4.

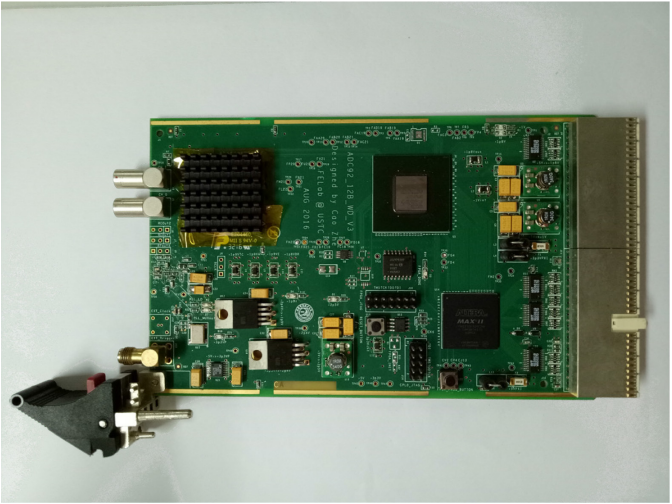

Fig. 4. The photo of the pre-amplifier.

### III. TEST RESULTS

In order to evaluate the performance of the instrument, a series of electronics tests has been carried out.

The test bench was installed with a vector signal generator R&S SMA 100A, an oscilloscope LeCroy WavePro 715 Zi, a group of band pass filters from 30.5 MHz to 498 MHz and test software in the PXI controller.

#### A. Mismatch error correction

In the time interleaved ADC mode, there are three mismatch errors between channels. The gain and offset mismatches and the timing skew among the sampling clock signals. may cause pattern noise and degrade the dynamic performance of the time interleaved ADC significantly.

In order to overcome the mismatch errors in the time interleaved ADC, many digital calibration techniques have been proposed. The core idea is to obtain the three kinds of mismatch parameters, and then to correct the gain and offset errors by addition and multiplication processes and complex algorithm for the time skew error. There are several methods for correcting the time skew error, such as the interpolation [15], the blind compensation [16], the method based on fractional delay filters [17] and the perfect reconstruction method [18].

In this project, a software based on the perfect reconstruction mismatch error correction algorithm is used to correct the mismatches offline.

#### B. Frequency response test

The frequency response test of the instrument is shown in fig. 5. The source of signal from the vector signal generator is calibrated by the oscilloscope first, and then sent to the instrument. The pre-amplifier module is set in a series of gain, so that the signal in different frequency is amplified or attenuated and then digitized by the waveform digitization module. At last, the digitized data transfer to the PXI controller and the amplitude of the signal is calculated by the test software.

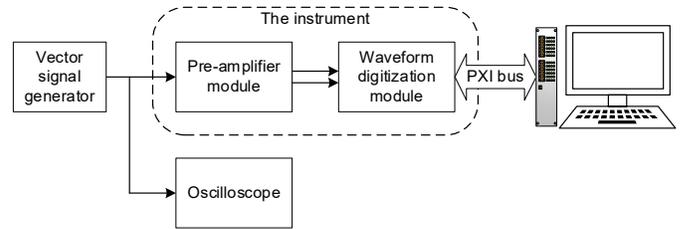

Fig. 5. The frequency response test bench. The source of signal from the vector signal generator is calibrated by the oscilloscope first, and then sent to the instrument to be digitized. The software in the PXI controller analysis the amplitude of the signal.

Frequency responses of the instrument in gain settings from -26.2 dB to 24.4 dB are depicted in Fig. 6. These curves demonstrate in a persuasive way that the system works well up to 450 MHz with moderate gain variation in different gain settings.

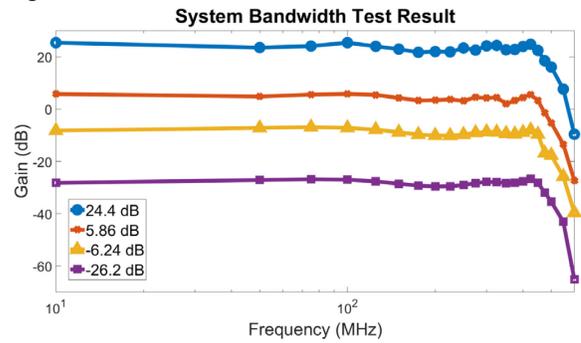

Fig. 6. Frequency responses of the system in different gain settings from maximum 24.4 dB to minimum -26.2 dB, blue line is 24.4 dB, red line is 5.86 dB, yellow line is -6.24 dB and purple line is -26.2 dB.

#### C. Dynamic performance test

According to IEEE Std. 1241-2010 [19], a series of tests to evaluate the performance of this high speed and high resolution ADC based instrument is conducted. As shown in Fig. 7, a high precision sine wave signal is generated by the vector signal generator and the band pass filter is employed to improve the performance of the sine wave by filtering the harmonics and out-of-band noise. Then the signal is digitized in different gain by the system. The software based on mismatch error correction algorithm analysis the dynamic feature, such as signal to noise ratio (SNR), total harmonic distribution (THD), (SFDR), signal to noise and distortion ratio (SINAD) and ENOB, via data from PXI bus.

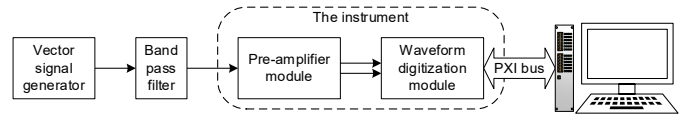

Fig. 7. The dynamic performance test bench.

By turning the frequency of input sine wave from 30.5 MHz to 498 MHz, systematic tests are conducted in a series of gain. The curves of different gain setting on SNR, SFDR, SINAD and ENOB are displayed in Fig. 8 to Fig. 12. The test results show that the instrument achieves an ENOB above 5.8 bits in the high gain set and above 7 bits in the low gain set in the bandwidth range.

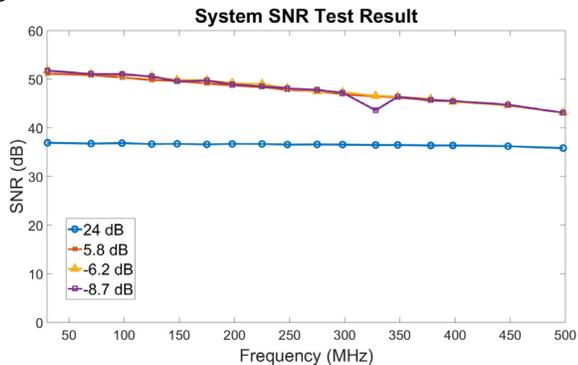

Fig. 8. SNR results of the system in different gain settings, blue line is 24 dB, red line is 5.8 dB, yellow line is -6.2 dB and purple line is -8.7 dB.

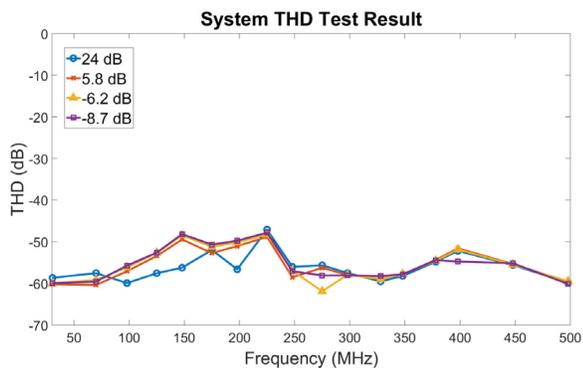

Fig. 9. THD results of the system in different gain settings, blue line is 24 dB, red line is 5.8 dB, yellow line is -6.2 dB and purple line is -8.7 dB.

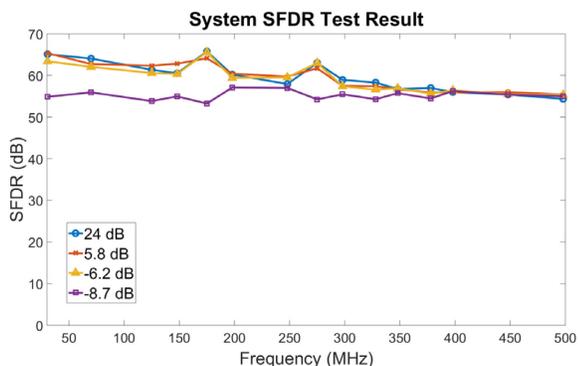

Fig. 10. SFDR results of the system in different gain settings, blue line is 24 dB, red line is 5.8 dB, yellow line is -6.2 dB and purple line is -8.7 dB.

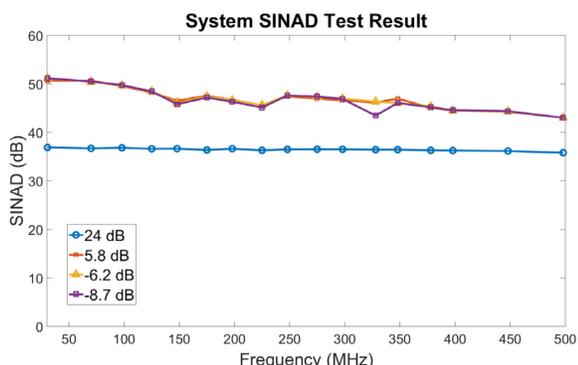

Fig. 11. SINAD results of the system in different gain settings, blue line is 24 dB, red line is 5.8 dB, yellow line is -6.2 dB and purple line is -8.7 dB.

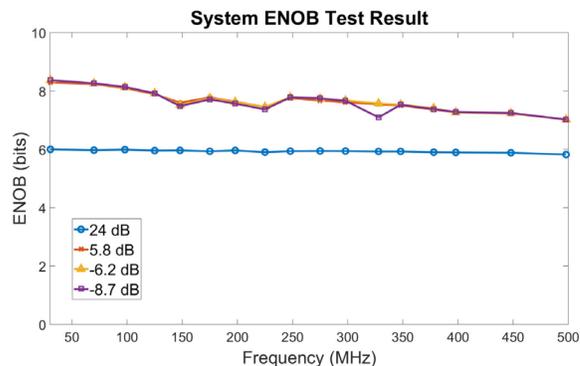

Fig. 12. ENOB results of the system in different gain settings, blue line is 24 dB, red line is 5.8 dB, yellow line is -6.2 dB and purple line is -8.7 dB.

IV. CONCLUSION

This paper presents a detector calibration instrument based on programmable gain waveform digitization. Two modules are designed for the system, including a pre-amplifier module based on programmable gain amplifier and attenuator and a PXI 3U waveform digitization module based on high speed and high-resolution ADC. This instrument has a large dynamic gain range from -26.2 dB to 24.4 dB and sample rate of 3.6 GSPS with 12-bit resolution. An electronics test platform is setup to evaluate the instrument and detailed tests have been done. The systematic measurement results reveal that the bandwidth of the module is from DC to 450 MHz and the ENOB is above 7 bits in high gain set and 5.8 bits in low gain set in the bandwidth range. According to the test results, the system has been proven to perform well. This variable gain amplifier and waveform digitization based instrumentation can bring better performance in various detector calibrations.